# Stability Analysis of the Linear Discrete Teleoperation Systems with Stochastic Sampling and Data Dropout

A.A. Ghavifekr[1], A.R. Ghiasi[*], M.A. Badamchizadeh[*], F. Hashemzadeh[*], P. Fiorini[**]

[1]Corresponding author at: Department of Electrical and Computer Engineering, University of Tabriz, Tabriz, Iran -Email: aa.ghavifekr@tabrizu.ac.ir
[*]Department of Electrical and Computer Engineering, University of Tabriz, Tabriz, Iran
Email: [agiasi, mbadamchi, hashemzadeh]@tabrizu.ac.ir
[**]Department of Computer Science, University of Verona, Verona, Italy
Email: paolo.fiorini@univr.it

*Abstract:* This paper addresses the stability conditions of the sampled-data teleoperation systems consisting continuous-time master, slave, operator, and environment with discrete-time controllers over general communication networks. The output signals of the slave and master robots are quantized with stochastic sampling periods which are modeled as being from a finite set. By applying an input-delay method, the probabilistic sampling system is converted into a continuous-time system including stochastic parameters in the system matrices. The main contribution of this paper is the derivation of the less conservative stability conditions for linear discrete teleoperation systems taking into account the challenges such as the stochastic sampling rate, constant time delay and the possibility of data packet dropout. The numbers of dropouts are driven by a finite state Markov chain. First, the problem of finding a lower bound on the maximum sampling period that preserves the stability is formulated. This problem is constructed as a convex optimization program in terms of linear matrix inequalities (LMI). Next, Lyapunov-Krasovskii based approaches are applied to propose sufficient conditions for stochastic and exponential stability of closed-loop sampled-data bilateral teleoperation system. The proposed criterion notifies the effect of sampling time on the stability-transparency trade-off and imposes bounds on the sampling time, control gains and the damping of robots. Neglecting this study undermines both the stability and transparency of teleoperation systems. Numerical simulation results are used to verify the proposed stability criteria and illustrate the effectiveness of the sampling architecture.

## 1. Introduction

During the past three decades, teleoperation systems have gradually grown through the world. They have been utilized in numerous applications such as mining vehicles and systems[1], underwater operation[2], telesurgery, minimally invasive surgery systems[3], space exploration[4], and many other applications in which human operators need to protect themselves from hazardous environments. In [5], a comprehensive survey of theories and developments in teleoperation systems is presented.

A typical teleoperation system is composed of human operator, master and slave robots, environment, and communication channel. Several teleoperation control architectures have been proposed in the literature[6]; including position-error-based (PEB), direct force reflection (DFR) and 4-channel structure[7]. Information signals are transmitted between master and slave robots through communication networks in the presence of unavoidable time delay. The communication delay can deteriorate the stability of the teleoperation system. Various continuous-time control schemes have been reported in literature to address this issue. The most widely used control methods include wave variables [8], adaptive control [9], and passivity theory [10]. Also, the input-to-state stability analysis of the general teleoperation system consisting of a network with stochastic time delays is investigated in [11].

Besides stability which is the fundamental requirement for every control system, the teleoperation system must be completely "transparent"[12] which means the slave robot accurately receives the master's commands and the master correctly feels the slave forces. Indeed, transparency is a measure of system's performance [13]. In this paper, tracking error between the master position/velocity and the slave position/velocity has been defined as a criterion for transparency of the teleoperation system.

The passivity is another important factor of teleoperation systems that has been proposed widely in the context of continuous-time teleoperation systems. This is due to well-known property of the passive systems which states that a feedback interconnection of passive systems is necessarily passive and stable[14]. Conventional methods for stability analysis of teleoperation systems assume that the human operator does not inject energy into the master robot and behaves in a passive manner.

Obviously, many studies concern the continuous-time control of teleoperation systems. Despite extensive studies considering continuous-time teleoperation systems, only a few papers have proposed analysis and controller design for discrete bilateral forms. Apparently, in today's digitalized world, it is of both practical significance and theoretical importance to analyze how a discrete control signal would influence behavior of a continuous-time dynamic network. This paper presents a method for the discrete-time teleoperation systems which can help to design controllers with high gains without losing stability and transparency.

The primary question is about necessity of using discrete-time controllers for teleoperation systems. In [15], analog and digital control of bilateral teleoperation systems have been compared comprehensively in theory and experiment. Constraints of analog controllers for these systems are highlighted, and guidelines are suggested to address them. By comparing hybrid parameters of continuous-time and discrete-time controlled bilateral teleoperation systems presented in a recent paper [16], it can be deduced that, besides all benefits of analog controllers, they cannot tackle problem of a discrete-time communication channel with unreliabilities such as packet-loss and data duplication or swapping.

Several control schemes have been proposed to study new challenges arising in discrete-time teleoperation systems, including step invariant transformation plus low pass filters [17], the Tustin method plus scattering operators [18], passive geometric method [19], and nonlinear control using input to state stability (ISS) [20]. One of the most important challenges in this area is that passivity of a discrete teleoperation system is not guaranteed due to energy leaks, which is caused by Zero Order Hold (ZOH). In [21], a ZOH energy-instilling effect has been investigated. It is represented that passivity of a teleoperation system can be jeopardized if the continuous-time controllers are substituted with discrete equivalents. Study done in [22], is one of the most primary studies that takes sampled-data controllers in the teleoperation systems into account. Both time delay and ZOH effects are considered in the stability analysis and a robust control structure is proposed for teleoperation systems. In [23], the novel passivity conditions are proposed for delay-free sampled-data teleoperation systems. Also, using the small gain theorem, and assuming that both transmitted position and force signals are subjected to time delay, absolute stability of the teleoperation systems is studied in [24]. Recent studies try to find passivity conditions for system components such as operator or the environment [25].

Using complex networked control systems in teleoperation systems have recently received considerable attention due to their benefits such as ease of maintenance, lower cost and considerable flexibility [26]. Network control systems may suffer from packet dropouts [27], time-varying transmission delays [28], communication constraints [29], and quantization errors [30]. These studies provide conditions for calculating the so-called Maximal Allowable Transmission Interval (MATI). The system remains stable as long as the transmission interval is smaller than the MATI [31, 32].

In network control systems, method of sampling can greatly affect performance of the system. Traditional studies mainly focus on the single-rate sampling, where all signals are sampled at one constant rate. In the multirate sampling [33, 34], different signals are sampled at various but constant rates. In the randomly sampled systems, the sampling rate can vary from one sample to another sample. This stochastic sampled-data method in network control systems has attracted attentions in recent years and numerous results have been reported in the literature [35-38]. In [39], it is declared that system performance can be improved if the sampling period varies via network condition. A crucial issue arises here is that variation of sampling periods may drastically deteriorate the stability of the controlled systems.

Stabilization of the sampled-data networked teleoperation systems is a challenging problem since imposing the stochastic sample and holds, time-delays and packet losses into the system might lead to its instability. Thus, stability of digitally implemented teleoperation systems needs further arguments. This paper is devoted to stochastic stability analysis of the sampled-data teleoperation systems.

A natural question arises: how to preserve stability of the discrete-time teleoperation systems when the networks are imposed by stochastic sampling, data packet loss, and time delays. Motivated by this, the main contributions of this paper are twofold:

*Calculating a lower bound for maximum allowable network-induced delay that preserves stability of the sampled-data teleoperation system.

*Providing conditions that guarantee stochastic stability of the bilateral teleoperation system in the presence of varying sampling periods and data dropout.

The first problem is constructed as a convex optimization program in terms of linear matrix inequalities (LMI), and Lyapunov-Krasovskii based approaches are applied to propose sufficient conditions for exponential stability of the closed-loop sampled-data teleoperation system. For the second problem, the numbers of data dropouts are driven by a finite state Markov chain. Using iterative approach, the time-varying sampling method is applied to model the network of teleoperation system, and the stochastic stability is satisfied by adding an extra LMI condition.

The rest of this paper is organized as follows: In section 2, modeling and preliminaries of teleoperation systems are formulated. In section 3, the proposed framework for sampled-data teleoperation systems with stochastic sampling is developed, and this model is used in section 4 to find a lower bound for maximum allowable sampling period to preserve stability. In section 5, stochastic stability of the system is discussed. Finally, in section 6, simulations are performed to demonstrate effectiveness of the main results.

*Notation:* Throughout the paper $R^n$ denotes the n-dimensional; Euclidean space. A matrix $P > 0 (P < 0)$ means $P$ is real symmetric and positive definite (negative definite). '$\|\,.\,\|$' represents the Euclidean norm of a vector or its corresponding induced norm for a matrix. Prob $\{\alpha|\beta\}$ means the occurrence probability of $\alpha$ conditional on $\beta$. For a symmetric matrix, $\lambda_{max}(A), \lambda_{min}(A)$ stand for the largest and smallest eigenvalues of $A$, respectively. $E(x|y)$ denotes the expectation of $x$ conditional on $y$.

For $-\tau \leq r \leq 0$, the absolutely continuous function $x_t$ is defined as $x_t(r) = x(t+r)$, and its norm is denoted by

$$\|x_t\|_\omega = \max|x_t(r)| + [\int_{-\tau}^{0}|x_t(r)|^2 dr]^{1/2}.$$

## 2. System Modelling and Preliminaries

In a bilateral teleoperation system, the operator sends commands to the slave robot using the master robot and the master receives position or force signals on the slave side. The general schematic of this system is illustrated in Fig. 1, where both robots and communication channel are integrated into a linear time invariant master-slave two-port network block. It is assumed that the master and slave robots have dynamics with the same structures.

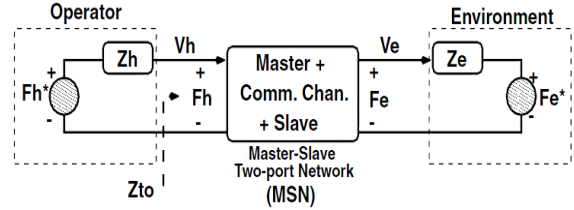

Fig. 1. Schematic of bilateral teleoperation network block diagram[40]

$V_h$ and $V_e$ are hand-master and slave-environment velocities, respectively. Also, $F_h, F_e$ denote the forces exerted by the operator's hand on the master and by the environment on the slave, respectively.

In the absence of friction, gravitational forces, and other disturbances, the dynamics of the master and the slave for ideal 2-DOF robots are given as follows:

$$m_m \ddot{q}_m + b_m \dot{q}_m = F_h - F_m \qquad (1)$$
$$m_s \ddot{q}_s + b_s \dot{q}_s = F_e - F_s$$

The notations $m$ and $s$ are used for the master and slave robots, respectively. Likewise, $m$ and $b$ denote mass and corresponding damping of robots. $q, \dot{q}$ are angular displacement and velocity signals. The LTI impedances of the operator and the environment are assumed to be passive, and indicated by $Z_h(s)$ and $Z_e(s)$, respectively.

$F_m$ and $F_s$ are control inputs in the master and slave sides. According to the general scheme that is represented in Fig. 1, the continues-time models of the operator and the environment are:

$$F_h = F_h^* - Z_h(s)sX_m \qquad (2)$$
$$F_e = F_e^* - Z_e(s)sX_s \qquad (3)$$

where "$s$" is a notation for the Laplace operator, and $X_m$ and $X_s$ indicate position of master and slave robots. Also, $F_h^*$ and $F_e^*$ are used for exogenous force inputs implied by the operator and the environment, respectively. The impedance modelling of the master and slave robots can be stated by

$$Z_m = \frac{1}{m_m s + b_m}$$
$$Z_s = \frac{1}{m_s s + b_s} \quad (4)$$

A hybrid matrix is a well-known method to model the position error based (PEB) teleoperators[41], where the slave robot follows the master robot position and vice versa. The basic definition is

$$\begin{bmatrix} F_h(s) \\ -sX_s(s) \end{bmatrix} = \begin{bmatrix} h_{11} & h_{12} \\ h_{21} & h_{22} \end{bmatrix} \begin{bmatrix} sX_m(s) \\ F_e(s) \end{bmatrix} \quad (5)$$

where

$$H(s) = \begin{bmatrix} Z_m + C_m \frac{Z_s}{Z_s + C_s} & \frac{C_m}{Z_s + C_s} \\ -\frac{C_s}{Z_s + C_s} & \frac{1}{Z_s + C_s} \end{bmatrix} \quad (6)$$

where $C_m$ and $C_s$ are controllers of the master and slave robots, respectively.
For any desired operator or environment dynamics, the ideal transparency can be defined as:

$$F_h = F_e \,,\, \dot{X}_m = \dot{X}_s \quad (7)$$

Thus, from (5) the ideal hybrid matrix can be expressed as:

$$H_{ideal} = \begin{bmatrix} 0 & 1 \\ -1 & 0 \end{bmatrix} \quad (8)$$

The aforementioned condition is satisfied if gains of the controllers are chosen large enough. However, as will be indicated in the next section, this will jeopardize stability of the system. Elements of the hybrid matrix have direct physical significance. $h_{11}$ is the impedance transmitted to the operator when the slave is in free space. The parameter $h_{21}$ represents velocity tracking fidelity when the slave robot is in free motion. $h_{22}, h_{21}$ are output admittance and force tracking fidelity when the master is in contact with a stiff hand.
To obtain the state-space form of (1) we define:

$$x_m = \begin{bmatrix} q_m \\ \dot{q}_m \end{bmatrix} \,,\, x_s = \begin{bmatrix} q_s \\ \dot{q}_s \end{bmatrix} \quad (9)$$

Then (1) can be written as follows:

$$\dot{x}_m = Ax_m + B(F_h - F_m)$$
$$\dot{x}_s = Ax_s + B(F_e - F_s) \quad (10)$$

where

$$A = \begin{bmatrix} 0 & 1 \\ 0 & -\frac{b}{m} \end{bmatrix} \,,\, B = \begin{bmatrix} 0 \\ \frac{1}{m} \end{bmatrix} \quad (11)$$

To facilitate the sampled-data controller design, first the discrete-time model of the aforementioned teleoperation system should be derived. Assuming the sampling period of $h_k$, the discrete equivalent counterpart for the master and slave robots are as follows:

$$x_m(k+1) = A_d x_m(k) + B_d F_m(k) + \int_{kh_k}^{(k+1)h_k} e^{[(k+1)h_k - \tau]A} B F_h(\tau) d\tau$$

$$x_s(k+1) = A_d x_s(k) + B_d F_s(k) \quad (12)$$

where

$$A_d = e^{Ah_k}, B_d = \int_{kh_k}^{(k+1)h_k} e^{[(k+1)h_k - \tau]A} B d\tau = \int_{0}^{h_k} e^{A\tau} B d\tau$$

**Remark 1.** It is notable that human and environment torques are still in the continuous-time form which is different compared to previous literature[42]. Since the torque from human and the torque resulting from contact with environment do not work in a discrete way, it is more reliable to treat them as continuous-time signals. Also, the delayed value of the $F_h$ can be measured by force/torque sensors. A more accurate analysis of the stability should consider the specific dynamics of human and environment. However, due to emphasis on stochastic stability in this paper, these dynamics are coupled with master and slave dynamics.

## 3. The proposed model for the sampled-data bilateral teleoperation system

In this section, the mathematical model of the discrete networked teleoperation system is established.
It is assumed that both position and velocity signals are measured at the sampling instants $\hat{t}_k$, $k \in N$. Each sampled signal is sent via network in one data packet. Sampled signals are transmitted via communication networks with constant time delay and data packet dropout. The possibility of data loss is modeled via switches in Fig. 2. The proposed model is composed of two time driven samplers with stochastic sampling periods and two event driven ZOHs. The general scheme is illustrated in Fig. 2, in which the discrete signals are represented by dash lines.

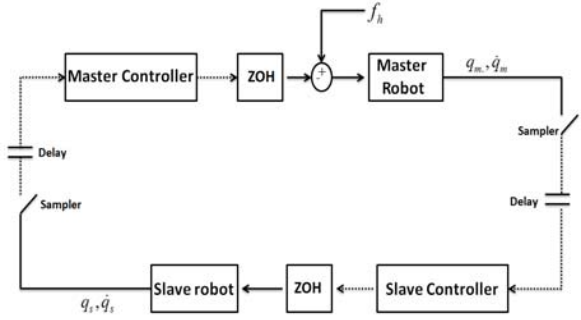

Fig.2. The proposed model for the sampled-data teleoperation system with stochastic sampling

The output signals of the slave and master robots sampled at $\hat{t}_k$. These signals are imposed by a constant time delay $T$ as they are transmitted through the network. It is notable that our proposed model allows the constant delay to be larger than the sampling interval $[\hat{t}_k, \hat{t}_{k+1}]$. Length of the $kth$ sampling period is defined by $h_k$, i.e.

$$h_k = \hat{t}_{k+1} - \hat{t}_k$$

**Assumption 1.** There exists $\varepsilon > 0$ such that $\hat{t}_{k+1} - \hat{t}_k > \varepsilon$. This states the fact that sampling processes cannot occur simultaneously in practice.

Both ZOHs are updated with new control signals at the instants $t_k$,

$$t_k = \hat{t}_k + T_k \quad k \in N \tag{13}$$

The control signals are kept constant by the event driven ZOH through the interval $[t_k, t_{k+1}]$. The elapsed time since the last sampling instant $\hat{t}_k$ is defined as:

$$\mu(t) \triangleq t - \hat{t}_k = t - t_k + T \tag{14}$$

$\mu(t)$ is known as the network-induced delay. We denote the largest amount of this parameter by $\gamma$, i.e.

$$\gamma = \sup(\mu(t)) = \sup(\hat{t}_{k+1} - \hat{t}_k) \tag{15}$$

Dividing $[\varepsilon, \gamma]$ into $n$ equal small intervals and considering $\tilde{k}$ as the instant that the control input reaches actuators, the following relation is proposed to calculate the next sampling instant:

$\hat{t}_{k+1} = \hat{t}_k + d_1$ for $\tilde{k} \in [d_1, d_2]$ and $\hat{t}_{k+1} = \hat{t}_k + \gamma$ for $\tilde{k} \geq \hat{t}_k + \gamma$.

where $d_1 = \gamma + a(\gamma - \varepsilon)/n, d_2 = \varepsilon + (a+1)(\gamma - \varepsilon)/n$ for $a = 0, 1, .., n-1$. Thus, the variable sampling periods switch in the finite set of $\chi = \{\varepsilon, (\gamma - \varepsilon)/n, ..., \gamma\}$.

In the proposed scheme, illustrated in Fig. 2, the P+d controller laws are utilized for both of the master and slave robots. A continuous-time form of this controller is designed in [43], in order to achieve passivity, bilateral force reflection, and master-slave coordination of the closed loop teleoperator.

The control signals can be redefined regarding the aforementioned equations. i.e.

$$\begin{aligned} F_m(t) &= -K_v(\dot{x}_m(\hat{t}_k) - \dot{x}_s(\hat{t}_k - T_2)) - \\ &\quad (K_d + P_\varepsilon)\dot{x}_m(\hat{t}_k) - K_P(x_m(\hat{t}_k) - x_s(\hat{t}_k - T_2)) \\ F_s(t) &= -K_v(\dot{x}_s(\hat{t}_k) - \dot{x}_m(\hat{t}_k - T_1)) - \\ &\quad (K_d + P_\varepsilon)\dot{x}_s(\hat{t}_k) - K_P(x_s(\hat{t}_k) - x_m(\hat{t}_k - T_1)) \end{aligned} \tag{16}$$

where $T_1, T_2 \geq 0$ are the forward and backward constant time delays. $K_p, K_v$ are the positive P+d control gains. $K_d$ is the dissipation gain to compensate the delayed P-control and $P_\varepsilon$ is a kind of additional damping for ensuring master-slave coordination. In [43], it is proved that by defining $K_d = \frac{T_1 + T_2}{2} K_p$, the required control objectives are satisfied. According to similar dynamics of the slave and master robots, it is natural to use the same controller gains for them.

## 4. Calculating maximum allowable sampling period to preserve stability of the system

In this section, sufficient stability conditions are provided for discrete-time teleoperation system presented in Fig. 2. A theorem is proposed in the form of LMIs to calculate a lower bound for the maximum network-induced delay that can preserve the exponential stability of the sampled-data teleoperation system.

Applying the input delay approach[44], (16) can be rewritten as:

$$\begin{aligned} F_m(t) &= -K_v(\dot{x}_m(t - \mu_s) - \dot{x}_s(t - \mu_s - T_2)) - \\ &\quad (K_d + P_\varepsilon)\dot{x}_m(t - \mu_s) - K_P(x_m(t - \mu_s) - x_s(t - \mu_s - T_2)) \\ F_s(t) &= -K_v(\dot{x}_2(t - \mu_s) - \dot{x}_m(t - \mu_s - T_1)) - \\ &\quad (K_d + P_\varepsilon)\dot{x}_s(t - \mu_s) - K_P(x_s(t - \mu_s) - x_m(t - \mu_s - T_1)) \end{aligned} \tag{17}$$

Considering (17), the following theorem is proposed in this paper to compute the maximum allowable sampling period that preserves stability of the teleoperation system described in (10).

**Theorem 1.** Consider the linear discrete teleoperation system defined in (12) and (17) with variable sampling intervals smaller than $\gamma$. For $\alpha > 0$, the system is uniformly exponentially stable, if symmetric positive definite matrices $X, P, R$ and $G$, with appropriate dimensions exist, satisfying

$$a: \Omega + \gamma \Lambda_1 < 0 \tag{18}$$

$$b: \begin{bmatrix} \Omega + \gamma \Lambda_2 & \gamma G \\ \gamma G^T & -\gamma e^{-\alpha \gamma} R \end{bmatrix} < 0$$

where

$$\Sigma = \begin{bmatrix} A_i & -B_i K_p & -B_i(K_v + K_d + P_\varepsilon) & -B_i & -B_i \end{bmatrix} \tag{19}$$

The subscript of i=m is used for the master robot and i=s is used for the slave robot. Also, we have:

$$\Omega = \Sigma^T [P \ 0] + [P \ 0]^T \Sigma + \alpha [I \ 0]^T P [I \ 0]$$

$$-[I \ -I]^T X [I \ -I] - \begin{bmatrix} I & -I \\ 0 & 0 \end{bmatrix}^T G^T - G \begin{bmatrix} I & -I \\ 0 & 0 \end{bmatrix}$$

$$\Lambda_1 = \begin{bmatrix} A_i & B_i K_p & B_i (K_v + K_d + P_\varepsilon) & -B_i & -B_i \\ 0 & I & 0 & 0 & 0 \end{bmatrix}^T R \times$$

$$\begin{bmatrix} A_i & B_i K_p & B_i (K_v + K_d + P_\varepsilon) & -B_i & -B_i \\ 0 & I & 0 & 0 & 0 \end{bmatrix} \quad (20)$$

$$+ \alpha [I \ -I]^T X [I \ -I] + \Sigma^T X [I \ -I] + [I \ -I]^T X \Sigma$$

$$\Lambda_2 = -\begin{bmatrix} 0 & 0 \\ 0 & I \end{bmatrix}^T G^T - G \begin{bmatrix} 0 & 0 \\ 0 & I \end{bmatrix}$$

**Proof.** The proof of this theorem is presented in appendix A. A modified Lyapunov Krasovskii Function (LKF)[31] is proposed for this proof.

This theorem provides sufficient stability conditions in the form of LMIs that can be solved accurately using available software. Using theorem 1, the problem of computing MASP can be formulated as an optimization program in terms of LMIs that is comparable with [44, 45].

## 5. Stochastic stability with data packet dropout and random sampling

In this section, the time-varying sampling method has been used to model networked control linear teleoperation system with finite data packet dropout by implying an iterative approach [46]. First, a definition for stochastic stability of teleoperation system with stochastic dynamics is expressed, and then another LMI condition is added to (18) to satisfy this definition.

Assume that $\bar{x}_i(k), \dot{\bar{x}}_i(k)$ (i=m for master and i=s for slave) denote position and velocity measurements that are successfully transmitted via the network, i.e. $\bar{x}_i(k) = x_{k-1}$.

**Assumption 2.** Successive transmission periods $\{d_j, j = 0,1,2,...\}$ are from the finite state of the Markov chain with transition probability $\tau_{\alpha\beta} = P\{d_{j+1} = \beta / d_j = \alpha\}, \forall \alpha, \beta \in S$ where $\sum_{\beta=1}^{M} \tau_{\alpha\beta} = 1$

We assume that the successive non-drop transmitted instants of $k$ are $0 = k_0 < k_1 < ... < k_j < ... < k_{max}$.

In order to describe the closed loop of the master and slave robots the following auxiliary term is defined:

$$\Upsilon(k_{j+1}) = \Phi_b^{k_{j+1}-k_j-1} \Phi_{h_{k_k}} \dot{x}_i(k_j) - (\Phi_b^{k_{j+1}-k_j-2} \Gamma_b + ... + \Gamma_b) K_v \dot{x}_i(k_j)$$

$$- \Phi_b^{k_{j+1}-k_j-1} \Gamma_{hk_j} K_v \dot{x}_i(k_{j-1}) + \Phi_b^{k_{j+1}-k_j-1} \Phi_{h_{k_k}} \dot{x}_r(k_j - T_n) +$$

$$(\Phi_b^{k_{j+1}-k_j-2} \Gamma_b + ... + \Gamma_b) K_v \dot{x}_r(k_j - T_n) +$$

$$\Phi_b^{k_{j+1}-k_j-1} \Gamma_{hk_j} K_v \dot{x}_r(k_{j-1} - T_n) + \Phi_b^{k_{j+1}-k_j-1} \Phi_{h_{k_k}} \dot{x}_i(k_j) -$$

$$(\Phi_b^{k_{j+1}-k_j-2} \Gamma_b + ... + \Gamma_b)(K_d + P_\varepsilon) \dot{x}_i(k_j)$$

$$- \Phi_b^{k_{j+1}-k_j-1} \Gamma_{hk_j} (K_d + P_\varepsilon) \dot{x}_i(k_{j-1}) + \Phi_b^{k_{j+1}-k_j-1} \Phi_{h_{k_k}} x_i(k_j) -$$

$$(\Phi_b^{k_{j+1}-k_j-2} \Gamma_b + ... + \Gamma_b) K_p x_i(k_j) - \Phi_b^{k_{j+1}-k_j-1} \Gamma_{hk_j} K_p x_i(k_{j-1}) +$$

$$\Phi_b^{k_{j+1}-k_j-1} \Phi_{h_{k_k}} x_r(k_j - T_n) + (\Phi_b^{k_{j+1}-k_j-2} \Gamma_b + ... + \Gamma_b) K_p x_r(k_j - T_n) +$$

$$\Phi_b^{k_{j+1}-k_j-1} \Gamma_{hk_j} K_p x_r(k_{j-1} - T_n) \quad (21)$$

where $d_{j+1} = k_{j+1} - k_j$, indicates the successive transmission period. It means that, the number of data packet dropout is $d_{j+1} - 1$. $i, r = m$ stand for the master robot and $i, r = s$ stand for the slave robot. $T_n$ for $n = 1$ and $n = 2$ is the forward and backward constant time delays, respectively.

By substituting $i = s, r = m$, and $n = 1$ in (21), the closed loop system of the slave robot can be described by following equation:

$$x_s(k_{j+1}) = \Upsilon(k_{j+1}) \quad (22)$$

Also, by substituting $i = m, r = s$, and $n = 2$ in (21), and applying the effect of the human operator, the closed loop system of the master robot can be described by the following equation:

$$x_m(k_{j+1}) = \Upsilon(k_{j+1}) + \int_{kh_k}^{(k+1)h_k} e^{[(k+1)h_k - \tau]A} BF_h(\tau) d\tau \quad (23)$$

(23) can be written as

$$x_m(k_{j+1}) = A(d_{j+1})\dot{x}_m(k_j) - B_1(d_{j+1}) K_v \dot{x}_m(k_j) -$$
$$B_2(d_{j+1}) K_v \dot{x}_m(k_{j-1}) + B_3(d_{j+1}) \dot{x}_s(k_j - T_2) +$$
$$B_4(d_{j+1}) K_v \dot{x}_s(k_j - T_2) + B_5(d_{j+1}) K_v \dot{x}_s(k_{j-1} - T_2) + B_6(d_{j+1}) \dot{x}_m(k_j)$$
$$- B_7(d_{j+1})(K_d + P_\varepsilon) \dot{x}_m(k_j) - B_8(d_{j+1})(K_d + P_\varepsilon) \dot{x}_m(k_{j-1}) -$$
$$B_9(d_{j+1}) x_m(k_j) - B_{10}(d_{j+1}) K_p x_m(k_j) - B_{11}(d_{j+1}) K_p x_m(k_{j-1}) +$$
$$B_{12}(d_{j+1}) x_s(k_j - T_2) + B_{13}(d_{j+1}) K_p x_s(k_j - T_2) +$$
$$B_{14}(d_{j+1}) K_p x_s(k_{j-1} - T_2) + \int_{kh_k}^{(k+1)h_k} e^{[(k+1)h_k - \tau]A} BF_h(\tau) d\tau \quad (24)$$

where

$$\Phi_{hk_j} = e^{Ah_{k_j}}, \Phi_b = e^{Ah_{max}}, \Gamma_{h_{kj}} = \int_0^{h_{kj}} e^{As} ds B, \Gamma_b = \int_0^{h_{k_{max}}} e^{As} ds B$$

Let

$$z(j) = [\dot{x}_m(k_j)^T, \dot{x}_m(k_{j-1})^T K_v^T, \dot{x}_s(k_j - T_2)^T, \dot{x}_s(k_{j-1} - T_2)^T K_v^T,$$
$$\dot{x}_m(k_{j-1})^T (K_d + P_\varepsilon)^T, x_m(k_j)^T, x_m(k_{j-1})^T K_p^T, x_s(k_j - T_2)^T,$$
$$x_s(k_{j-1} - T_2)^T K_p^T]^T \quad (25)$$

be the augmented state vector. The compact form of dynamics can be written as:

$$z(j+1) = \tilde{A}(d_{j+1})z(j) \tag{26}$$

where

$$\tilde{A}(d_{j+1}) = E(d_{j+1}) + F(d_{j+1})\bar{K} \tag{27}$$

where

$$\bar{K} = [K_v \ K_d \ K_P \ 0] \tag{28}$$

The same equations can be written for the slave side.
$M-1$ is considered as the upper bound of the data packet dropout, and $d_j = \alpha \in S = \{1,2,...,M\}$.

**Definition. 1[47]:** A system with stochastic dynamics such as (26) is stochastic stable, if for initial distribution $d_0 \in S$, and every initial state $z_0 = z(0)$, there exists a finite positive matrix $Q$ such that

$$E(\sum_{j=0}^{\infty}\|z(j)\|^2 \mid d_0) < z_0^T Q z(0) \tag{29}$$

Now, according to the aforementioned definition, another LMI inequality is required to satisfy (29)[37].
Assuming $X(1) > 0, X(2) > 0,..., X(M) > 0, G > 0$, are positive symmetric matrices and matrix $Y$ and $G - X(\alpha)$, $\forall \alpha \in S$ are full rank, it is shown that holding the following LMI leads to a stochastic stable teleoperation system.

$$\begin{pmatrix} X(\alpha) - G - G^T & \Psi^T \\ \Psi & -\Sigma \end{pmatrix} < 0 \tag{30}$$

where

$$\Psi = [\sqrt{r_{\alpha 1}} G^T E^T(1) + Y^T F^T(1))......\sqrt{r_{\alpha M}}(G^T E^T(M) + Y^T F^T(M))]^T$$

$$\Sigma = diag\{X(1),...,X(M)\}$$

$$G = \begin{pmatrix} G_{11} & G_{12} \\ G_{21} & G_{22} \end{pmatrix}$$

To this end, $V(z(j),d_j) = z^T(j)W(d_j)z(j)$ is added as a stochastic term to the proposed Lyapunov function in Appendix A, where $W(\alpha) = X^{-1}(\alpha), d_j = \alpha \in S$.
Thus,

$$E\{[V(z(j+1),d_{j+1}) - V(z(j),d_j)] | d_j = \alpha\} = z^T(j)\Omega z(j)$$

where

$$\Omega = \sum_{\beta=1}^{M} r_{\alpha\beta} \tilde{A}^T(\beta)W(\beta)\tilde{A}(\beta) - W(\alpha) \tag{31}$$

$G - X(\alpha)$ is full rank and we can write

$$(G - X(\alpha))^T X^{-1}(\alpha)(G - X(\alpha)) = G^T X^{-1}(\alpha)G - G - G^T + X(\alpha) > 0 \tag{32}$$

Then we have:

$$X(\alpha) - G - G^T > -G^T X^{-1}(\alpha)G.$$

Thus, LMI (30) implies that $\Omega < 0$ and we have

$$E\{[V(z(j+1),d_{j+1}) - V(z(j),d_j)]\} \leq -\lambda_{\min}(-\Omega)\|z(j)\|^2 \tag{33}$$

For any $N \geq 1$, it can be rewritten as:

$$E\{[V(z(N+1),d_{N+1}) - V(z(0),d_0)]\} < -\inf\{\lambda_{\min}(-\Omega)\}E\{\sum_{j=0}^{N}\|z(j)\|^2\} \tag{34}$$

It can be concluded that:

$$E\{\sum_{j=0}^{N}\|z(j)\|^2\} \leq \frac{1}{\inf\{\lambda_{\min}(-\Omega)\}}[E\{V(z(0),d_0)\}] \tag{35}$$

Assuming $N \to \infty$ and $Q = (\frac{1}{\inf\{\lambda_{\min}(-\Omega)\}})I$, then

$$E\{\sum_{j=0}^{\infty}\|z(j)\|^2\} \leq \frac{1}{\inf\{\lambda_{\min}(-\Omega)\}}[E\{V(z(0),d_0)\}] = z^T(0)Qz(0)$$

Which is satisfied Eq. (29) in definition 1.

## 6. Numerical Simulation

In this simulation, pair of 2-serial-links revolute-joint robots are considered. The system parameters are chosen to be the same as those in [42]. i.e., $M = 8.4796 \times 10^{-3} kgm^2$, $b = 114.6 \times 10^{-6} Ns/m$. After discretization, the state space matrices will be

$$A_d = \begin{bmatrix} 1 & 9.999932 \times 10^{-4} \\ 0 & 0.999986 \end{bmatrix}, B_d = \begin{bmatrix} 5.896478 \times 10^{-5} \\ 0.117929 \end{bmatrix}.$$

Human operator is modeled as a PD position tracking controller with its spring and damping gains as $75N/m$ and $50Ns/m$. The contact with the environment occurs when the slave robot reaches $4rad$.
We choose the P-action gain $K_p$ to be $50N/m$. Delays $T_1, T_2$ are assumed to be 0.6s and 0.4s, respectively.
The dissipation gain $K_d = 25N/m$ is chosen according to time delays. The extra damping $P_\varepsilon$ and the D-action gain $K_v$ are also set to be $0.001K_d$ and 1.
It is assumed that sampling period is time-variant and varies randomly with a specific probability among three values. Suppose the minimum sampling period of the sensor is 0.045s. However, the maximum sampling period should be found by solving LMIs in (18). By applying condition (18), it is calculated as 0.21s. Indeed, it is the lower bound of the maximum allowable sampling period that preserves the exponential stability of the system. If the sampling period exceeds this amount, the system may be unstable. Fig. 3 presents this case for 0.3s. This computed lower bound for the maximum allowable sampling period is compared with the proposed LMIs in [44, 45]. The results are presented in Table 1. According to this table, the results of this paper compare favorably with previous studies and are less conservative.

Table 1. Comparison of the computed maximum allowable sampling period that preserves exponential stability

|  | [44] | [45] | Theorem 1 |
|---|---|---|---|
| MASP | 0.108s | 0.14s | 0.21s |

The feasible values of the sampling periods are considered $h_1 = 0.045\sec, h_2 = 0.09\sec, h_3 = 0.21\sec$.

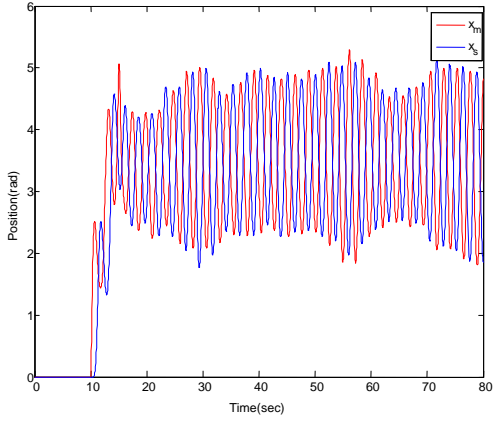

Fig. 3. Trajectories of the master and slave robots for sampling time of 0.3s that leads to instability of the system

Suppose that maximum transmission period in both slave and master sides is $M = 3$ and $\{d_j = 0,1,...\}$ is a Markov chain with transition probability described by

$$R = \begin{pmatrix} 0.12 & 0.52 & 0.36 \\ 0.11 & 0.80 & 0.09 \\ 0.53 & 0.14 & 0.33 \end{pmatrix} \quad (36)$$

Solving the linear matrix inequality (30), we obtain the solution given by

$$X(1) = \begin{pmatrix} 12.35 & -17.84 & -5.6 \\ -17.84 & 33.24 & 7.89 \\ -5.6 & 7.89 & 41.5 \end{pmatrix}$$

$$X(2) = \begin{pmatrix} 16.47 & -12.36 & -5.21 \\ -12.36 & 23.57 & 1.78 \\ -5.21 & 1.75 & 41.5 \end{pmatrix}$$

$$X(3) = \begin{pmatrix} 15.36 & -19.41 & -4.87 \\ -19.41 & 29.87 & 3.52 \\ -4.87 & 3.52 & 53.14 \end{pmatrix}$$

$$G = \begin{pmatrix} 19.72 & -17.25 & -6.23 \\ -17.25 & 39.44 & 1.15 \\ -6.23 & 1.15 & 85.21 \end{pmatrix}$$

According to definition 1, it is known that bilateral system is stochastic stable. Trajectories of the master and slave robots and tracking error are illustrated in Fig. 4 and Fig. 5, respectively.

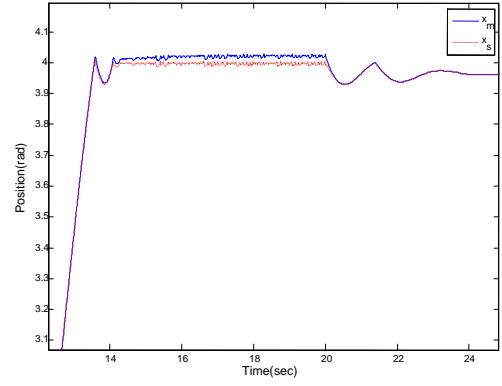

Fig. 4. Trajectories of the master and slave robots for stochastic sampling time in the presence of packet loss

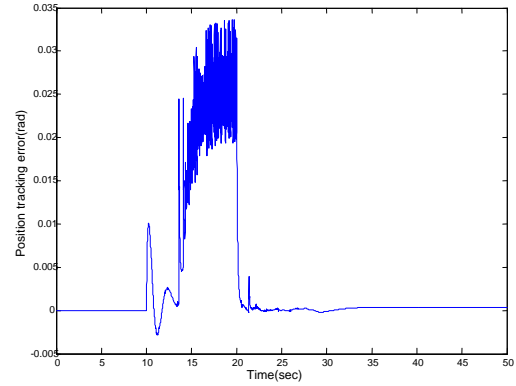

Fig. 5. Position tracking error signal between the master and the slave

If the constant sampling period is adopted, $h_3$ should be chosen as the sampling period to avoid network congestion. It is common when the network is occupied by several users. Evaluation of position signals for $h_3$ is depicted in Fig. 6.

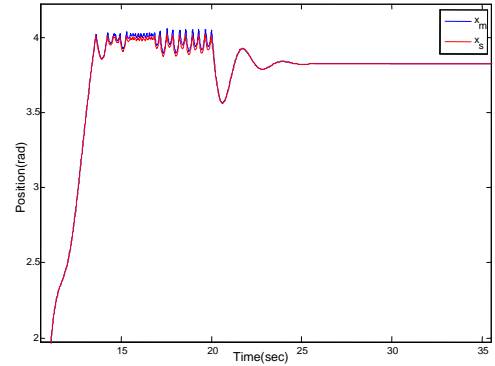

Fig. 6. Trajectories of the master and slave robots for the maximum allowable sampling period

In Fig. 7, the trajectory of the slave robot is illustrated for different values of MASP which were presented in table 1.

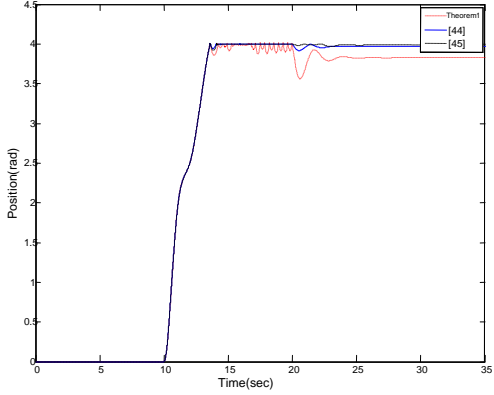

Fig. 7. Trajectories of the slave robot for different MASP in table 1

It is demonstrated that we can preserve the stability of the system with the higher maximum allowable sampling period and with less conservative conditions.

For teleoperation systems, in addition to stability, transparency is also crucial. To improve transparency, control gains should be increased. However, this may cause an undesirable effect on stability of the system.

One of the advantages of stochastic sampling is that we can make a tradeoff between stability and transparency. This is illustrated in Fig.8. In this case, we increase controller gains as $K_P = 100, K_d = 50$ and consider the sampling periods as $h_1 = 0.0225s, h_2 = 0.045s, h_3 = 0.09s$.

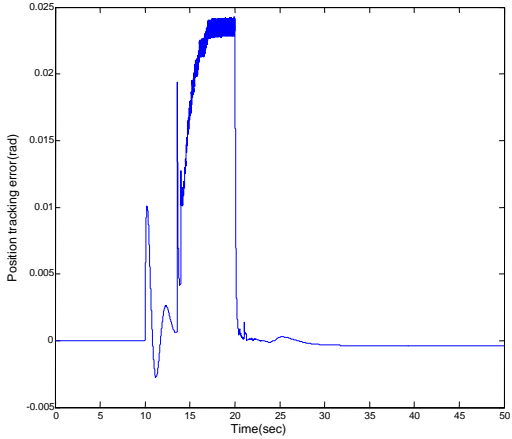

Fig. 8. Position tracking error signal between the master and the slave for reduced sampling time

Regarding to the above simulation result, better tracking performance has been achieved. Indeed, by considering stochastic features in the data packet losses and calculating maximum allowable sampling reasonable tradeoff can be considered between stability and transparency.

## 7. Conclusion

Although bilateral control of continuous-time teleoperation systems has been studied relatively well in previous studies, there are a few materials about sampled-data controlled teleoperation systems in the control literature. Teleoperation systems with discrete structure have been proven to be difficult in terms of transparency, stability, and implementation.

The proposed stability conditions in this paper imposed bounds on the sampling time. A lower bound for the maximum allowable network-induced delay is calculated that preserves stability of the sampled-data teleoperation system. Also, LMI conditions to guarantee the stochastic stability of the discrete bilateral teleoperation system in the presence of varying sampling periods and data dropout are provided. Comparing with other schemes, this method imposes less conservative conditions on the sampling time. These analyses can help design guidelines to have better transparency and stable teleoperation systems.

For future works on the sampled-data bilateral teleoperation systems, stability and transparency conditions should be extended for other structures, and 4-channel control architecture should be taken into account. In addition, the variable time delay between the master and slave can be considered.

## Appendix A. Proof of Theorem 1.

This appendix provides the mathematical proof of the theorem 1, presented in section 4.

First, we extract the following lemma [44]:

**Lemma 1.** For a constant $\lambda > 0$, $x(t)$ that is the solution of $\dot{x} = f(t, x_t)$ is uniformly exponentially stable if there exists a Lyapunov function $V(t, x_t)$, satisfying

$$a : c_1 |x_t(0)|^2 \leq V(t, x_t) \leq c_2 \|x_t\|_\omega^2$$
$$b : \dot{V}(t, x_t) + \lambda V(t, x_t) < 0 \quad \forall t \neq t_n, n \in N \quad (A.1)$$
$$c : 0 < \varepsilon < t_{n+q} - t_n$$

where $c_1, c_2$ and $\varepsilon$ are positive scalars.

Now for the proof, we indicate that the LMIs in Theorem 1 are fulfilled the conditions of the aforementioned lemma. The dynamics of the master robot can be rewritten as

$$\dot{x} = f(t, x_t) \to f(t, x_t) = Ax(t) + BKx_t(-\mu) \quad (A.2)$$

$$\dot{x}_m(t) = \begin{bmatrix} A_m & -B_m K_p & -B_m(K_v + K_d + P_\varepsilon) & -B_m & -B_m \end{bmatrix} \chi_m(t)$$

where

$$\chi_m(t) = \begin{bmatrix} x_1(t) & x_1(\hat{t}_k) & \dot{x}_1(\hat{t}_k) & \dot{x}_2(\hat{t}_k - T_2) & x_2(\hat{t}_k - T_2) \end{bmatrix}^T \quad (A.3)$$

and for the slave robot

$$\dot{x}_s(t) = \begin{bmatrix} A_s & -B_s K_p & -B_s(K_v + K_d + P_\varepsilon) & -B_s & -B_s \end{bmatrix} \chi_s(t) \quad (A.4)$$

where

$$\chi_s(t) = \begin{bmatrix} x_2(t) & x_2(\hat{t}_k) & \dot{x}_2(\hat{t}_k) & \dot{x}_1(\hat{t}_k - T_1) & x_1(\hat{t}_k - T_1) \end{bmatrix}^T \quad (A.5)$$

Since the samplers and controller gains of robots are assumed to be equal, the Lyapunov function's validity just is indicated for the master side and slave equations are omitted owing to the lack of space. The same results are obtained by substituting (A.4) and (A.5) in the proposed function for the slave side. Let a candidate Lyapunov Krasovskii Function (LKF) be defined as:

$$V(t,x_t) = x(t)^T P x(t) + (\gamma - \mu)(x(t) - x(t_n))^T X (x(t) - x(t_n)) +$$

$$(\gamma - \mu) \int_{t-\mu}^{t} e^{\alpha(s-t)} \left[ \dot{x}^T(s) \; x^T(t_n) \right] R \left[ \dot{x}^T(s) \; x^T(t_n) \right]^T ds \quad (A.6)$$

where $P, R$ and $X$ are positive definite matrices. In order to satisfy the first condition of (A.1), lower and upper bounds should be computed for the proposed LKF. According to the quadratic function properties, it can be shown that:

$$V_{\min} < V(t,x_t) < V_{\max} \quad (A.7)$$

where

$$V_{\min} = \lambda_{\min}(P) |x(t)|^2$$

$$V_{\max} = \lambda_{\max}(P) |x(t)|^2 + \gamma \lambda_{\max}(R)(1+\gamma) \|x_t\|_\omega^2 + 4\gamma \lambda_{\max}(X) \|x_t\|_\omega^2$$

Thus, the LKF candidate satisfies inequality (A.1-a). Next, the derivative of the proposed LKF is calculated to study (A.1-b). To this end, assuming $h(t) = G^T \chi_m(t)$ where $G$ is a matrix with proper size, $\dot{V}$ can be calculated as

$$\dot{V} + \alpha V = \dot{V}_1 + \dot{V}_2 + \dot{V}_3 + \alpha(V_1 + V_2 + V_3) \leq$$

$$\dot{x}_m^T P x_m + x_m^T P \dot{x}_m + \mu h^T e^{\alpha \gamma} R^{-1} h - [x_m^T - x_m^T(t_n) \;\; \mu x_m^T(t_n)] h$$

$$- h^T [x_m^T - x_m^T(t_n) \;\; \mu x_m^T(t_n)]^T$$

$$+ (\gamma - \mu) \left[ \dot{x}_m^T \; x_m^T(t_n) \right] R \left[ \dot{x}_m^T \; x_m^T(t_n) \right]^T - \alpha V_2$$

$$- (x_m(t) - x_m(t_n))^T X (x_m(t) - x_m(t_n)) +$$

$$(\gamma - \mu)(\dot{x}_m^T(t) X (x_m(t) - x_m(t_n)) + (x_m(t) - x_m(t_n))^T X \dot{x}_m(t))$$

$$+ \alpha V_2 + \alpha x_m(t)^T P x_m(t)$$

$$+ \alpha(\gamma - \mu)(x_m(t) - x_m(t_n))^T X (x_m(t) - x_m(t_n))$$

That can be rewritten as:

$$\dot{V} + \alpha V \leq$$

$$\chi_m^T \left[ A_m \;\; -B_m K_p \;\; -B_m(K_v + K_d + P_\varepsilon) \;\; -B_m \;\; -B_m \right]^T P [I \; 0]$$

$$+ [I \; 0]^T P \left[ A_m \;\; -B_m K_p \;\; -B_m(K_v + K_d + P_\varepsilon) \;\; -B_m \;\; -B_m \right]$$

$$+ \alpha [I \; 0]^T P [I \; 0]$$

$$+ \mu G e^{\alpha \gamma} R^{-1} G^T - \begin{bmatrix} I & -I \\ 0 & \mu I \end{bmatrix}^T G^T - G \begin{bmatrix} I & -I \\ 0 & \mu I \end{bmatrix} + \ldots \quad (A.8)$$

$$+ (\gamma - \mu) \begin{bmatrix} A_m & B_m K_p & B_m(K_v + K_d + P_\varepsilon) & -B_m & -B_m \\ 0 & I & 0 & 0 & 0 \end{bmatrix}^T \times$$

$$R \begin{bmatrix} A_m & B_m K_p & B_m(K_v + K_d + P_\varepsilon) & -B_m & -B_m \\ 0 & I & 0 & 0 & 0 \end{bmatrix}$$

$$+ \alpha [(\gamma - \mu) - 1][I \; -I]^T X [I - I]$$

$$+ (\gamma - \mu) \left[ A_m \;\; -B_m K_p \;\; -B_m(K_v + K_d + P_\varepsilon) \;\; -B_m \;\; -B_m \right]^T \times$$

$$X [I - I] + (\gamma - \mu)[I \; -I]^T X \times$$

$$\left[ A_m \;\; -B_m K_p \;\; -B_m(K_v + K_d + P_\varepsilon) \;\; -B_m \;\; -B_m \right] \chi_m$$

It is notable that for $\mu = 0$, LMI (18-a) satisfies $\dot{V} + \alpha V < 0$. Also, using the Schur complement, LMI (18-b) implies $\dot{V} + \alpha V < 0$ for $\mu = \gamma$. Thus, it can be concluded that LMIs (18) are sufficient conditions to satisfy $\dot{V} + \alpha V < 0$ for any $\mu \in (0, \gamma)$. Furthermore, according to the assumption 1, inequality (A.1-c) assumed to be a true statement for all of the successive sampling instants. This completes the proof. ∎